\documentclass[prc,twocolumn,showpacs,amsmath,amssymb]{revtex4}
\usepackage{times}
\usepackage{graphicx}
\usepackage{dcolumn}
\usepackage{bm}
\usepackage{amstext}

\begin{document}

\title{Internal stucture of clusters in $^{112-122}$Ba nuclei within relativistic mean field theory}
\author{M. Bhuyan$^{1,2}$, S.K. Patra$^1$, P. Arumugam$^3$, and Raj K. Gupta$^4$}
\affiliation{$^1$Institute of Physics, Sachivalaya Marg, Bhubaneswar-751 005, India, \\
$^2$School of Physics, Sambalpur University, Jyotivihar, Burla-768 019, India,\\
$^3$Department of Physics, Indian Institute of Technology, Roorkee-247 667, India,\\
$^4$Department of Physics, Panjab University, Chandigarh-160 014, India.
}
\date{\today}

\begin{abstract}

We study the clustering structure and the internal or sub-structure of clusters in $^{112-122}$Ba nuclei within the 
framework of relativistic mean field theory in an axially deformed cylindrical co-ordinate.  We calculate the total 
density distribution, and the individual neutrons and protons density distributions. From the analysis of the 
clustering confugurations of the density distributions of various shapes, we find different sub-structures inside the 
Ba nuclei considered here. The important step, carried out for the first time, is the counting of number of protons and 
neutrons present in the clustering region(s). $^{12}$C is shown to consitute the cluster configuration of Ba nuclei in
most cases, with $^{2,3}$H and $^4$He constituting the neck between two fissioning symmetrical fragments.

\end{abstract}
\pacs{21.10.Dr, 21.60.-n, 21.60.Jz, 21.60.Gx, 24.10.Jv, 23.70.+j}
\maketitle

\section{Introduction}

Instabilities of ''stable" nuclei against exotic cluster decays were first studied by one of us (RKG) and collaborators 
\cite{gupta93,kumar94,kumar95,kumar96} for the neutron-deficient $^{108-116}$Xe, $^{112-120}$Ba, $^{116-124}$Ce, 
$^{120-124}$Nd, $^{124-128}$Sm, and $^{128-132}$Gd and neutron-rich $^{146}$Ba, $^{152}$Ce, $^{156}$Nd, $^{160}$Sm, and
$^{164}$Gd nuclei, on the basis of the preformed cluster model (PCM) of Malik and Gupta \cite{gupta88,malik89} (see, also 
the works of other authors \cite{poenaru91,poenaru95}). The $N=Z$, $\alpha$-nuclei  $^{8}$Be, $^{12}$C, $^{16}$O, $^{20}$Ne, 
$^{24}$Mg, and $^{28}$Si are predicted to be the most probable emitters from neutron-deficient parents, with a much 
smaller decay half-life compared to $N>Z$ clusters predicted preferably or observed from neutron-rich parents (e.g., 
$^{14}$C from $^{146}$Ba or $^{222}$Ra with $^{132}$Sn or $^{208}$Pb as daughter, respectively). Thus, other than $\alpha$ 
particle, $^{12}$C decay of $^{112-120}$Ba nuclei is shown to be the most probable one with $^{100}$Sn and its heavier 
isotopes as the daughter nuclei. The ground-state decay of Ba, however, could not be established as yet 
\cite{oganess94,gugliel95}, and a new phenomenon of intermediate mass fragments (IMFs) with 3$\le$Z$\le$9, also referred 
to as `clusters' or `complex fragments', is observed to be emitted from the compound nucleus $^{116}$Ba$^*$ formed in 
$^{58}$Ni+$^{58}$Ni$\rightarrow ^{116}$Ba$^*$ reactions at both the high \cite{campo88,campo91} and medium energies 
\cite{campo98,commara2k}. The measured IMF cross-section $\sigma_{IMF}$ for the $^{116}$Ba$^*$ decay at all the above 
mentioned medium and high energies are so far understood only on the preformed-clusters (PCM) based dynamical cluster-decay 
model (DCM) of one of us (RKG) and collaborators \cite{gupta06}. The DCM describes the $^{116}$Ba$^*$ data on $\sigma_{IMF}$ 
reasonably well, and predicts an additional fusion-fission of $^{116}$Ba$^*$ which consists of fragments at the heavy end 
of symmetric and near symmetric division (14$\le$Z$\le$28), very recently observed at GANIL \cite{bonnet08} for the decays 
of $^{118,122}$Ba$^{*}$ nuclei produced in $^{78,82}$Kr+$^{40}$Ca reactions at a lower incident energy, and also explained 
more recently \cite{kumar09} on DCM. It should thus be interesting to see if mean-field approaches, such as the relativistic 
mean field model, support such a clustering structure for the ground and/ or excited states of Ba nuclei.

Relativistic mean field (RMF) calculations for light nuclei \cite{arum05} have very successfully shown the 
${\alpha}-{\alpha}$ clustering in $^8$Be, a benchmark nucleus, and the $\alpha$-clustering and halo structures of 
(${\alpha}+{\alpha}$)-core or (${\alpha}+{\alpha}$+p)-core plus $x$n for the stable and exotic Be and B nuclei. For the 
N=Z, $\alpha$-nuclei $^{12}$C and $^{16}$O, the RMF formalism gives the several known ground-state (g.s.) structures such 
as the 3$\alpha$-equilateral triangle (co-existing with spherical shape) and 4$\alpha$-tetrahedron or kite-like, plus the 
3$\alpha$- and 4$\alpha$-linear chains for their excited states. Also, both $\alpha$- and non-$\alpha$-cluster structures, 
like $^{10}$B+$^{10}$B, $^{12}$C+$^{12}$C, $^{12}$C+$\alpha$+$^{12}$C and $^{16}$O+$^{16}$O, and 5-$\alpha$-trigonal 
bipyramid and pentagonal bipyramid (hollow sphere) are obtained for $^{20}$Ne to $^{32}$S nuclei, but {\it no} 5-$\alpha$, 
6-$\alpha$ chains, etc., are seen. Furthermore, $^{56}$Ni shows \cite{gupta08a} a preferred N=Z, $\alpha$-nucleus 
clustering for states with deformations up to hyper-deformation ($\beta\le$2.45). Similarly, for heavy actinides 
($^{222}$Ra, $^{232}$U, $^{236}$Pu and $^{242}$Cm), the RMF gives \cite{patra07} not only the N$\approx$ Z, $\alpha$-like 
clustering in the g.s. but also the exotic N$\neq$Z clustering in excited states. Signatures of clustering structure in RMF 
calculations of super-heavy Z=114 and 120, N=172-184 nuclei \cite{patra07,shar06} are also obtained in terms of exotic 
N$\neq$Z clusters at the centre of their super-deformed g.s. or the clustering in to two large and some small pieces is 
universal for all super-deformed ground states in Z=120 nuclei. The super-superheavy $^{238}$U+$^{238}$U$\rightarrow 
^{476}$184$^*$ nucleus also supports the clustering phenomenon, with a kind of triple fission of an exotic cluster in the 
neck region of two equal fragments of N=Z matter \cite{gupta07}. The actual internal or sub-structure structure of the 
clusters, however, was not determined in these calculations, which is one of the aims of the present study.  

Taking RMF model as an established tool for the cluster structure in nuclei, in this paper we demonstrate its application
to medium-heavy $^{112-122}$Ba nuclei, with an additional attempt to determine for the first time the internal structure(s) 
of cluster(s), i.e., the number of protons and neutrons in a cluster. The paper is so designed that Section II describes 
the relativistic mean field theory, and Section III gives the results obtained from our calculation. Finally a brief 
summary and concluding remarks are given in Section IV.

\section{The Relativistic Mean Field (RMF) Method}

The relativistic Lagrangian density for a nucleon-meson many-body system \cite{sero86,ring90}
\begin{eqnarray}
{\cal L}&=&\overline{\psi_{i}}\{i\gamma^{\mu}
\partial_{\mu}-M\}\psi_{i}
+{\frac12}\partial^{\mu}\sigma\partial_{\mu}\sigma
-{\frac12}m_{\sigma}^{2}\sigma^{2}\nonumber\\
&& -{\frac13}g_{2}\sigma^{3} -{\frac14}g_{3}\sigma^{4}
-g_{s}\overline{\psi_{i}}\psi_{i}\sigma-{\frac14}\Omega^{\mu\nu}
\Omega_{\mu\nu}\nonumber\\
&&+{\frac12}m_{w}^{2}V^{\mu}V_{\mu}
+{\frac14}c_{3}(V_{\mu}V^{\mu})^{2} -g_{w}\overline\psi_{i}
\gamma^{\mu}\psi_{i}
V_{\mu}\nonumber\\
&&-{\frac14}\vec{B}^{\mu\nu}.\vec{B}_{\mu\nu}+{\frac12}m_{\rho}^{2}{\vec
R^{\mu}} .{\vec{R}_{\mu}}
-g_{\rho}\overline\psi_{i}\gamma^{\mu}\vec{\tau}\psi_{i}.\vec
{R^{\mu}}\nonumber\\
&&-{\frac14}F^{\mu\nu}F_{\mu\nu}-e\overline\psi_{i}
\gamma^{\mu}\frac{\left(1-\tau_{3i}\right)}{2}\psi_{i}A_{\mu}.
\end{eqnarray}

From this Lagrangian we obtain the field equations for the nucleons and mesons. These equations are solved by expanding 
the upper and lower components of the Dirac spinors and the boson fields in an axially deformed harmonic oscillator basis 
with an initial deformation $\beta_{0}$. The set of coupled equations is solved numerically by a self-consistent iteration 
method. The centre-of-mass motion (c.m.) energy correction is estimated by the usual harmonic oscillator formula 
$E_{c.m.}=\frac{3}{4}(41A^{-1/3})$. The quadrupole deformation parameter $\beta_2$ is evaluated from the resulting proton 
and neutron quadrupole moments, as $Q=Q_n+Q_p=\sqrt{\frac{16\pi}5} (\frac3{4\pi} AR^2\beta_2)$. The root mean square (rms) 
matter radius is defined as $\langle r_m^2\rangle={1\over{A}}\int\rho(r_{\perp},z) r^2d\tau$; here $A$ is the mass number, 
and $\rho(r_{\perp},z)$ is the deformed density. The total binding energy and other observables are also obtained by using 
the standard relations, given in \cite{ring90}. We use the well known NL3 parameter set \cite{lala97}. This set reproduces 
the properties of stable nuclei. As outputs, we obtain different potentials, densities, single-particle energy levels, 
radii, quadrupole deformations and the binding energies. For a given nucleus, the maximum binding energy corresponds to 
the ground state and other solutions are obtained as various excited intrinsic states.

The density distribution of nucleons play the prominent role for studying the internal structure of a nucleus. For a
different quadrupole deformation, the density distribution $\rho(r_{\perp}, z)$ inside the nucleus must vary. For example, 
the $\rho(r_{\perp}, z)$ for a spherical nucleus is symmetrical in $(\rho, z)-$plane. However, it is highly asymmetric for 
a largely deformed nucleus. Knowing the density distribution of the spherical or (oblate/ prolate) deformed confuguration,
we can calculate the number of nucleons for each confuguration, defined as 
\begin{equation}
n=\int_{z_{1}}^{z_{2}}\int_{r_{1}}^{r_{2}}\rho(r_{\perp},z) d\tau,
\end{equation}
with $n$ as number of neutrons N, protons Z or mass A. Though a straight forward calculation, this is being carried out 
here for the first time.

\section{Calculations and Results}

\subsection{Ground state properties of Ba nuclei using the RMF formalism}

\subsubsection {Potential Energy Surface}
 
We first calculate the potential energy surface (PES) by using the RMF formalism in a constrained calculation 
\cite{patra09,flocard73,koepf88,reinhard89,hirata88}, i.e., instead of minimizing the $H_0$, we have minimized 
$H'=H_0-\lambda Q_{2}$, with $\lambda$ as a Lagrange multiplier and $Q_2$, the quadrupole moment. The term $H_0$ is the 
Dirac mean field Hamiltonian for RMF model (the notations are standard and its form can be seen in Refs. 
\cite{ring90,koepf88,hirata88}). In other words, we get the constrained binding energy from 
$BE_c=\sum_{ij}\frac{<\psi_i|H_0-\lambda Q_2|\psi_j>}{<\psi_i|\psi_j>}$ and the free energy from  
$BE=\sum_{ij}\frac{<\psi_i|H_0|\psi_j>}{<\psi_i|\psi_j>}$. The converged free energy solution does not depend on the 
initial guess value of the basis deformation $\beta_0$ as long as it is nearer to the minimum in PES. However, it converges 
to some other local minimum when $\beta_0$ is drastically different, and in this way we evaluate a different intrinsic 
isomeric state for a given nucleus.

\begin{figure}[ht]
\vspace{0.7cm}
\begin{center}
\includegraphics[width=0.95\columnwidth]{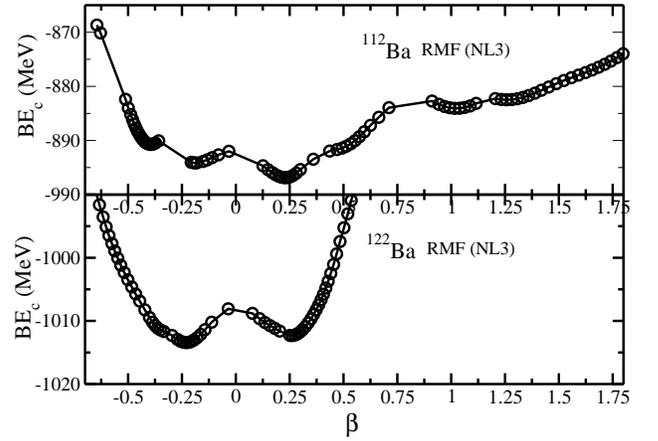}
\caption{The potential energy surfaces for $^{112}$Ba and $^{122}$Ba nuclei, i.e, constrained binding energy ($BE_c$) as a 
function of quadrupole deformation parameter in RMF(NL3) calculations. }
\end{center}
\label{Fig.1}
\end{figure}

\begin{table*}[t]
\caption{The RMF(NL3) results of binding energy, the quadrupole deformation parameter $\beta_{2}$, and the root mean square 
radii of charge $r_{c}$ and matter $r_{m}$ for the $^{112-122}$Ba nuclei. The energy is in MeV, and radii in fm. The
experimental data are from \cite{audi03,moha08}, and the binding energies marked with star (*) are the extrapolated 
values. The finite range droplet model (FRDM) binding energy and $\beta_2$ are taken from \cite{moller95} and the
theoretical charge radius $r_c$ is from Hartree-Fock-BCS (HFBCS) method \cite{goriely01}. No experimental data are available for $r_c$.
}
\begin{tabular}{|c|c|c|c|c|c|c|c|c|c|c|c|c|c|c|c|c|}
\hline
Nucleus &\multicolumn{2}{|c|}{BE}&\multicolumn{2}{|c|}{$r_{rms}$: RMF(NL3)}&\multicolumn{2}{|c|}{$\beta_{2}$}&BE&$r_c$&$\beta_2$ \\\hline
& RMF(NL3) &Expt.& $r_{m}$&$r_{c}$& RMF(NL3) &Expt.&FRDM&HFBCS&FRDM \\ \hline
$^{112}$Ba & 860.79 & & 4.80 & 4.99 &  -0.39 &&&&\\
           & 893.50 & & 4.65 & 4.79 &  -0.17 & &&&\\
           & 895.38 & & 4.62 & 4.74 &  0.24 &&894.89&4.72&0.21 \\
           & 882.36 & & 5.35 & 5.48 &  1.24 & &&&\\
           & 897.33 & & 12.17 & 12.21 &  10.71 &&&& \\
$^{114}$Ba & 918.10 & & 4.78 & 4.90 &  -0.39 & &&&\\
           & 920.13 & 922.26 & 4.65 & 4.75 &  0.24&&921.27&4.74&0.24\\ 
           & 909.35 & & 5.33 & 5.45 &  1.19 & &&&\\
           & 920.61 & & 12.16 & 12.20 &  10.58 & &&&\\
$^{116}$Ba & 943.66 & 947.024$^*$ & 4.80 & 4.90 & -0.39 &&&& \\
           & 945.70 & & 4.71 & 4.79 & 0.36 && 946.85&4.78&0.28 \\
           & 934.15 & & 5.36 & 5.46 &  1.20 & &&&\\
           & 942.38 & & 12.13 & 12.16 & 10.40 & &&&\\
$^{118}$Ba & 969.16 & & 4.73 & 4.81 &  -0.24 &&&& \\
           & 971.38 & 971.022$^*$& 4.75 & 4.82 &  0.33 &&970.75&4.80&0.29 \\
           & 962.45 & & 12.11 & 12.13 & 10.24 & &&&\\
$^{120}$Ba & 991.64 & & 4.75 & 4.81 & -0.23 &&&& \\
           & 993.94 &993.600 & 4.77 & 4.83 &  0.32 &&993.44&4.82&0.28 \\
           & 975.04 & & 12.55 & 12.58 & 10.97 & &&&\\
$^{122}$Ba & 1013.53 &  & 4.77 & 4.82 &  -0.22 &&&& \\ 
           & 1015.52 & 1015.499& 4.80 & 4.84 &  0.32 & 0.354&1015.21&4.81&0.27 \\
           & 999.40 & & 12.09 & 12.10 &  9.95 & &&&\\
\hline
\end{tabular}
\end{table*}

The PES for the representative $^{112}$Ba and $^{122}$Ba nuclei are shown in Fig. 1 for a wide range of $\beta_2$ starting 
from oblate to prolate deformation. The upper pannel is for $^{112}$Ba and the lower one for $^{122}$Ba. For $^{112}$Ba, 
we notice the minima are around $\beta_2$=-0.39, -0.20, 0.23, 1.02 and 1.20, corresponding to binding energy $BE_c$= 890.6, 
894.2, 896.7, 884.2 and 882.7 MeV, respectively, and their energy differences between the nearest consecutive minima are 
3.60, 2.50, 12.93 and 1.52 MeV. However, in case of $ ^{122}$Ba, only two minima exist around $\beta_2$=-0.23 and 0.26 
at $BE_c$= 1013.4 and 1012.4 MeV. The intrinsic energy difference between these two configurations is 1.03 MeV. From this 
figure, it is clear that there exists two identical major minima at $\beta_2\approx$-0.23 and 0.23 in both the $^{112}$Ba 
and $^{122}$Ba nuclei. A further investigation of the diagram shows that actually the multi-minima structure of $^{112}$Ba 
disappears gradually with the increase of mass number in the isotopic chain of Ba, and reaches to only two minima 
configuration, one oblate and another prolate, at mass number A=122.

\subsubsection{Binding Energies and Quadrupole Deformation Parameter}

The binding energies and quadrupole deformation parameters $\beta_2$ for $^{112-122}$Ba isotopes are evaluated for the 
ground as well as intrinsic excited states. The obtained results are tabulated in Table I, together with the experimental 
or extrapolated values, wherever available. The experimental value of g.s. $\beta_2$ is available for $^{122}$Ba 
only. Since the neutron-deficient $^{112-120}$Ba isotopes lie near the proton drip-line, their deformation parameters are 
not yet known. Table I shows that both the g.s. binding energies,
charge radius ($r_c$)  and $\beta_2$ values agree well with the experimental 
data and with the theoretical calculations \cite{moller95,goriely01}. 

As discussed above, lighter Ba isotopes have several intrinsic minima, where each minimum corresponds to a deformation and 
a binding energy. The largest binding energy minimum corresponds to the ground state and all other minima are the excited 
intrinsic states. In this way, the ground state $\beta_2$ is 0.24 for $^{112}$Ba. Similarly, the g.s. deformations
for $^{114}$Ba, $^{116}$Ba, $^{118}$Ba, $^{120}$Ba and $^{122}$Ba are 0.24, 0.36, 0.33, 0.32 and 0.32, respectively. All 
other intrinsic state deformations are also listed in Table I.  For $^{112,114,116}$Ba nuclei we get a solution at a highly 
deformed configuration of $\beta_2\sim 1.2$, whereas this hyper-deformed minimum is washed out with increase of mass number 
in the Ba isotopic chain. This means there is no hyper-deformed solutions 
for $^{118-122}$Ba (see also, Fig. 4
where such highly deformed configurations are shown only for $^{112,114,116}$Ba). If the nucleus is further deformed 
($\beta_2\sim$10), it breaks into two fragments (see also Fig. 5) with a separation radius $r_m$ of about 12 fm. An 
analysis of this structure shows two distinct (fission-like) fragments connected with a thin density distribution of 
nucleons, which can be considered like a very thin neck. We find that the composition of the (connecting) neck region is 
an isotopic chain of hydrogen or helium nuclei $^{2,3}$H and $^{4}$He (discussed below).
  
\subsubsection{The root-mean-square radii $r_{rms}$}

In this subsection, we discuss our calculation on root mean square (rms) radii of matter distribution, with various 
quadrupole deformations from oblate to prolate and then from super-deformation to hyper-deformation. The matter rms radius 
$r_m$ and charge distribution rms radius $r_c$ are listed in Table I for different $\beta_2$ values. From Table I, it is 
observed that the $r_m$ increases with increase of quadrupole deformation. Finally, both $r_m$ and  $r_c$ show a large 
extension of about 12 fm at $\beta_2\sim$10.5. As already pointed out above, at this point of deformation, the two 
fragments get separated from each other leaving a thin density distribution of matter in between the two fragments or 
clusters.

\begin{figure}[h]
\vspace{0.3cm}
\begin{center}
\includegraphics[width=0.75\columnwidth]{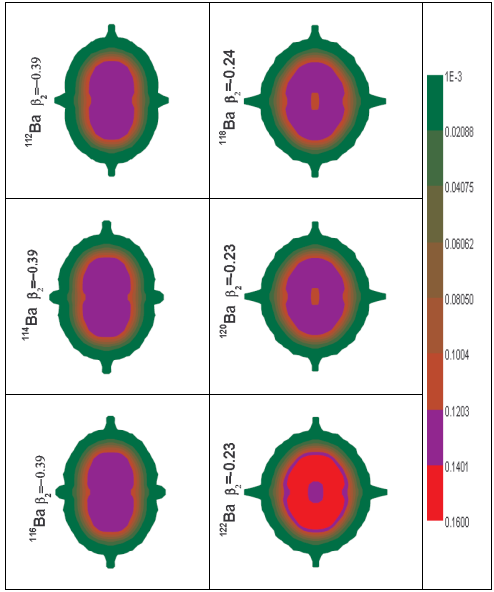}
\caption{(Color on line) The cluster confugurations of $^{112-122}$Ba for the oblate states in RMF(NL3).
}
\end{center}
\label{Fig.2}
\end{figure}

\begin{figure}[h]
\begin{center}
\includegraphics[width=0.75\columnwidth]{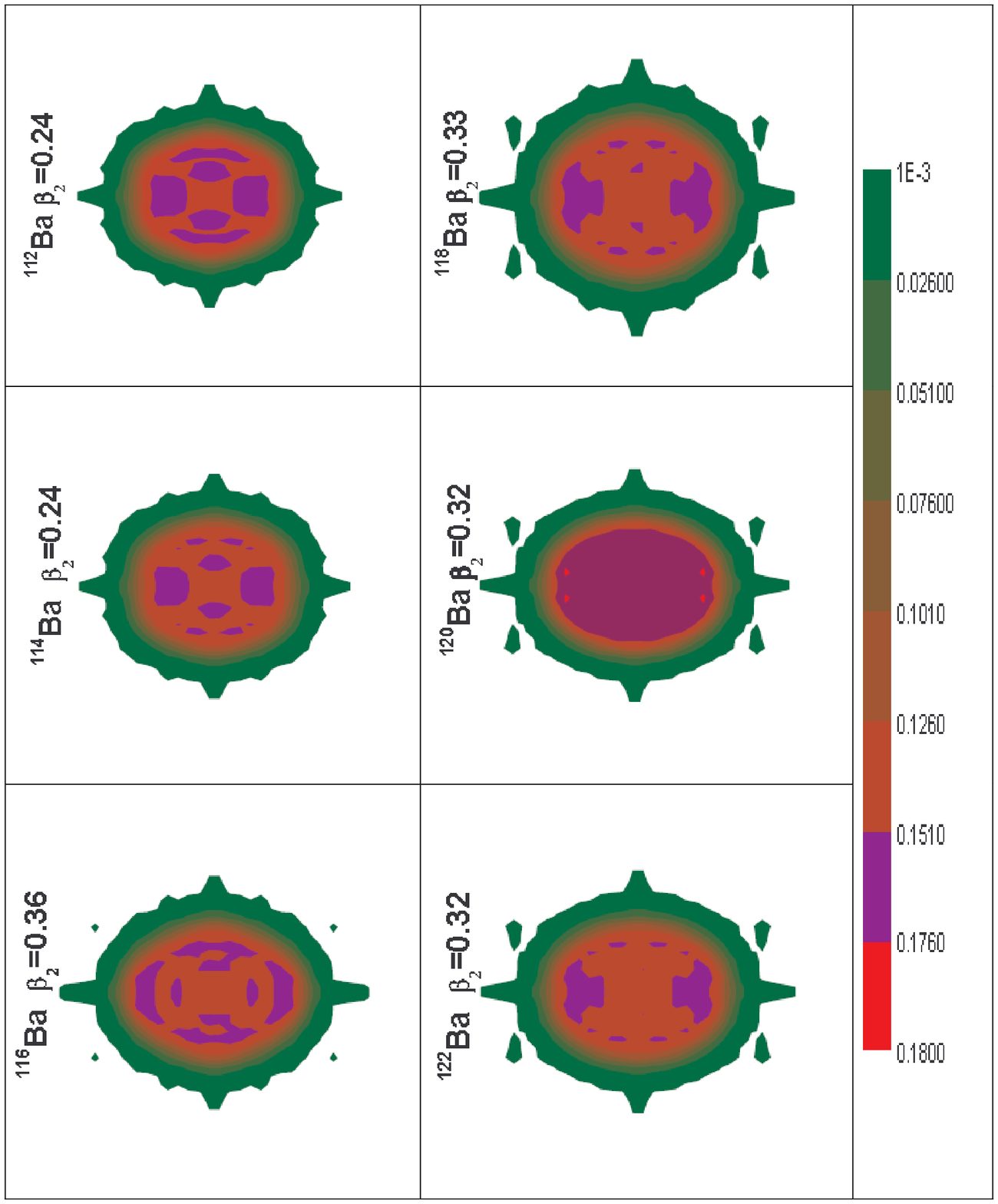}
\caption{(Color on line) The cluster confugurations of $^{112-122}$Ba for the ground states in RMF(NL3).
}
\end{center}
\label{Fig.3}
\end{figure}

\begin{figure}[h]
\vspace{0.3cm}
\begin{center}
\includegraphics[width=0.49\columnwidth]{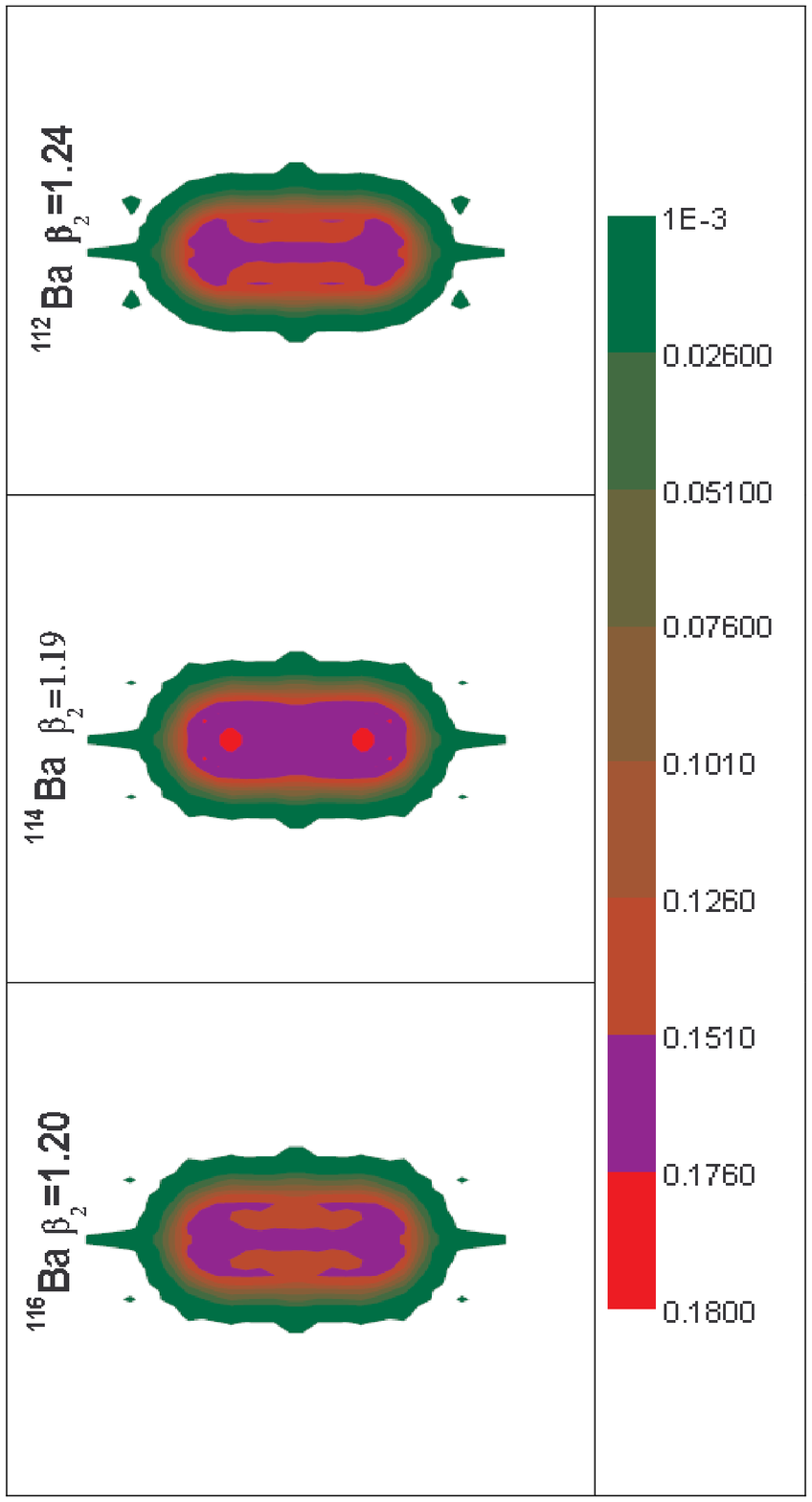}
\caption{(Color on line) The cluster confugurations of $^{112-122}$Ba for the prolate states in RMF(NL3).
}
\end{center}
\label{Fig.4}
\end{figure}

\begin{figure}[h]
\begin{center}
\includegraphics[width=0.75\columnwidth]{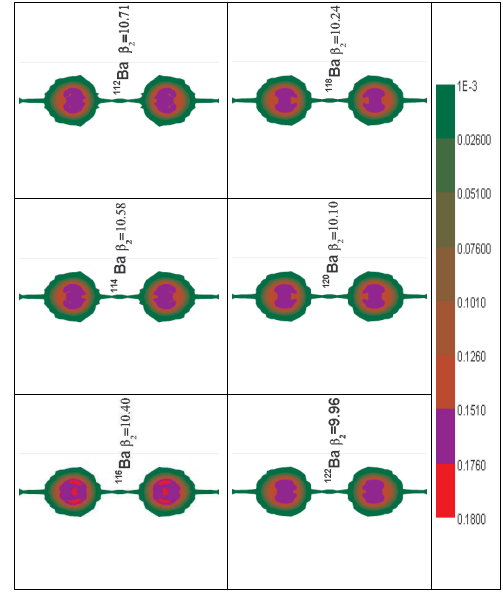}
\caption{(Color on line) The cluster confugurations of $^{112-122}$Ba for the hyper-deformed states in RMF(NL3).
}
\end{center}
\label{Fig.5}
\end{figure}

\subsection {Cluster confugurations in $^{112-122}$Ba nuclei}

Figures 2 to 5 show the density plots for all the possible solutions, starting from oblate to prolate deformations of the 
Ba isotopic chain. Fig. 2 gives the density distributions for $^{112-122}$Ba for the oblate state solutions around the 
deformation parameter $\beta_{2}\sim$-0.35, and Figs. 3, 4 and 5 show, respectively, the prolate state solutions  of 
$^{112-122}$Ba around $\beta_{2}\sim$0.25, 1.1 and 10.5. From these graphs the clustering structures in these nuclei are 
quite evident. Considering the colour code, starting from deep black with maximum to white (or light gray) bearing the 
minimum density, one can analyse the distribution of nucleons inside the various isotopes at various shapes. In Fig. 2 the 
minimum density for the oblate-state starts from 0.001 fm$^{-3}$ and goes up to maximum 0.16 fm$^{-3}$, but in case of the 
prolate-states in Figs. 3, 4 and 5, it starts from 0.001 fm$^{-3}$ and goes up to 0.18 fm$^{-3}$, which shows that the size 
of cluster nucleus is smaller in the oblate-state than that of the prolate-state. A careful inspection of the density 
distributions in different regions of the nucleus clearly shows the formation of various cluster(s) inside the nuclei, 
which are listed in Table II as clusters I, II or III for a given deformation $\beta_2$.  

Further, if we increase the value of deformation parameter $\beta_{2}$ to $\sim$10.4-10.7 or more, the nucleus gets 
separated into two fragments with a neck formation as shown in Fig. 5 for the isotopic chain of $^{112-122}$Ba nuclei. 
Another prominent observation is that there is no configuration of deformation parameter $\beta_{2}\sim$1.2 in case of 
$^{118-122}$Ba, similar to what is shown in Fig. 4 for lighter mass Ba isotopes. If we compare the cluster confugurations 
of each nucleus, it is clear that there is a gradual change in the confuguration inside the nucleus, i.e., the clusters 
inside the nucleus for different $\beta_{2}$ values are distinct from each other, and for each isotope of $^{112-122}$Ba 
chain.

\begin{table}[th]
\caption{The number of nucleons $A_{clus.}$, the protons $Z_{clus.}$, and neutrons $N_{clus.}$ in a cluster inside the 
$^{112-122}$Ba nuclei for different solutions of deformations $\beta_{2}$ obtained from the RMF(NL3) formalism.
The range of the cluster ($r_1,r_2;z_1,z_2)$ are in fm.
}
\begin{tabular}{|c|c|c|c|c|c|c|}
\hline
Nucleus & $\beta_{2}$ & Cluster No.& $A_{clus.}$ & $Z_{clus.}$ & $N_{clus.}$ &Cluster\\ 
&  & range ($r_1,r_2; z_1,z_2$)& & & &\\ \hline

$^{112}$Ba & -0.39 & I (-0.7,0.7;-3.0,-1.7) & 11.6 & 5.6 & 6.0 & $^{12}$C\\
           & 0.24 & I (1.9,4.5;-1.5,1.5 & 36.0 & 17.7 & 18.3 & $^{36}$Ar\\
           & & II (-1.3,1.3;-2.6,-1.2)& 13.0 & 6.3 & 6.7 & $^{13}$C\\
           & 10.71 & I (-6.3,6.3;-3.5,3.5) & 1.6 & 0.7 & 0.8 & $^{2}$H\\
$^{114}$Ba & 0.24 & I (2.1,4.6;-1.5,1.5) & 34.7 & 16.9 & 17.9 & $^{35}$Cl\\
           & & II (-1.1,1.1;1.2,2.5) & 12.6 & 6.1 & 6.4 & $^{13}$C\\
           & & III (0.7,2.0;3.3,3.7) & 2.3 & 1.1 & 1.2 & $^{2}$H\\
           & 1.19 & I (3.9,5.4;-0.7,0.7) & 8.4 & 4.1 & 4.3 & $^{8}$Be\\
           & 10.58 & I (-6.6,6.3;-3.8,3.8) & 2.2 & 1.1 & 1.1 & $^{2}$H\\
$^{116}$Ba & 0.36 & I (4.4,6.2;-1.6,1.6) & 25.3 & 12.3 & 13.1 & $^{25}$Mg\\
           & & II (2.6,3.4;-0.9,0.9) & 6.1 & 2.8 & 3.4 & $^{6}$Li\\
           & 10.40 & I (11.0,12.2;-0.8,0.8) & 7.5 & 3.7 & 3.8 & $^{8}$Be\\
           & & II (9.9,12.8;2.3,2.9) & 6.6 & 3.2 & 3.4 & $^{6}$Li\\
           & & III (-6.1,6.1;-3.4,3.4) & 1.9 & 0.9 & 1.0 & $^{2}$H\\
$^{118}$Ba & -0.24 & I (-0.6,0.6;-1.2,1.2) & 12.85 & 6.9 & 5.9 & $^{13}$C\\
           & 0.33 & I (2.5,5.5;-1.6,1.6) & 42.0 & 19.9 & 22.2 & $^{42}$Ca\\
           & & II (0.7,2.0;3.3,3.6) & 1.7 & 0.8 & 0.9 & $^{2}$H\\
           & & III (-0.4,0.4;-2.1,-1.6) & 1.6 & 0.7 & 0.8 & $^{2}$H\\
           & 10.24 & I (-6.2,6.2;-3.7,3.7) & 2.8 & 1.2 & 1.6 & $^{3}$H\\
$^{120}$Ba & -0.23 & I (-0.6,0.6;-1.2,1.2) & 12.9 & 5.8 & 7.1 & $^{13}$C\\
           & 0.32 & I (-5.5,-5.0;-0.7,-1.2) & 0.9 & 0.3 & 0.6 & $^{1}$H\\
           & 10.97 & I (-6.3,6.3;-4.3,4.3) & 3.8 & 1.6 & 2.1 & $^{4}$He\\
$^{122}$Ba & -0.22 & I (-1.0,1.0;-1.5,1.5) & 24.4 & 11.2 & 13.2 & $^{23}$Na\\
           & 0.32 & I (2.5,5.3;-1.7,1.7) & 42.6 & 19.7 & 22.8 & $^{43}$Ca\\
           & & II (-2.0,-0.8;3.4,3.7) & 1.6 & 0.7 & 0.8 & $^{2}$H\\
           & 9.96 & I -6.1,6.1;-3.7,3.7) & 2.4 & 1.0 & 1.4 & $^{2}$H\\
\hline
\end{tabular}
\end{table}

\subsection{Counting of nucleons in clusters formed inside the $^{112-122}$Ba nuclei}
   
In this subsection, we count the number of nucleons in different clusters formed inside the $^{112-122}$Ba isotopic chain, 
listed in the Table II. The density distributions of these clusters, obtained from the RMF(NL3) formalism for different 
solutions of deformation parameters $\beta_{2}$ from oblate to prolate confugurations, are already shown above in Figs. 2 
to 5. From these distribution plots of each oblate or prolate configuration, we find the number of nucleons by using the 
general formula in Eq. (2), for not only the total nucleons (using total density distribution of the nucleus) but also for 
protons and neutrons (using individual density distributions), which is listed in the Table II. For this calculation, we 
first find the range of the integral, i.e., the lower and the upper limits of the axes (r and z) from the plots in Figs. 2 
to 5, and then evaluate the integral for each case. 

Table II first shows that the number of protons and neutrons from the individual density distributins gives the total 
number of nucleons which is calculated here from the total density distribution of the nucleus for each isotope of Ba with
different $\beta_2$ solutions. Secondly, the internal cluster configuration of $^{112-122}$Ba nuclei contain $^{12}$C or 
$^{13}$C as the most common cluster. For the oblate-state ($\beta_{2\sim}$-0.35, Fig. 2), the density distribution is 
uniform for $^{114,116}$Ba isotopes, but in cases of $^{112,118,120}$Ba isotopes we find a $^{12}$C or $^{13}$C cluster, 
and for $^{122}$Ba a $^{23}$Na cluster. The ground-state density distributions (around $\beta_{2}\sim$0.25, Fig. 3) of 
$^{112,114}$Ba contain $^{13}C$ cluster, but $^{116-122}$Ba contain clusters of some other elements like $^{6}$Li and 
$^{1,2}$H. The density distributions for the prolate-state ($\beta_{2}\sim$1.2, Fig. 4) are again uniform for the isotopes 
$^{112,116}$Ba, but show $^{8}$Be cluster confuguration for $^{114}$Ba. Note that there are no solutions for 
$\beta_{2}\sim$ 1.2 for the isotopic chain $^{118-122}$Ba. For the very large $\beta_{2}\sim$10.5, the internal 
confugurations of the isotopic chain $^{112-122}$Ba are of the form of two separated (identical) nuclei which are connected 
by a neck like confuguration. The neck confugurations contain simply the hydrogen isotopes $^{2,3}$H or $^4$He nucleus. As 
already noted in the Introduction, the existence of $^{12}$C cluster inside the Ba nuclei has been of interest both from 
experimental and theoretical points of views. The important point is that $^{12}$C cluster is formed inside the Ba nuclei, 
and are not from the neck region where $^{2,3}$H or $^4$He nuclei are shown to exist. In other words, $^{12}$C constitutes 
the cluster structure of Ba nuclei.

\section{Summary and Conclusions}

Concluding, we have calculated the gross nuclear properties and the nucleon density distributions for the isotopic chain 
$^{112-122}$Ba using the deformed relativistic mean field (RMF) formalism with NL3 parameter set. The gross properties, 
like the binding energy, deformation parameter $\beta_{2}$ and the charge radius $r_{c}$, show qualitative similarity 
between the experimental and RMF calculated values. Analysing the nuclear density distributions, we get the internal or 
sub-structure of clusters in Ba isotopes which we find to consist mostly of $^{12,13}$C and several other clusters like 
$^{1,2}$H, $^6$Li and $^8$Be. Some heavier clusters of Na, Cl, Mg, Ar and Ca are also obtained. With the increase of 
deformation ($\beta_{2}\sim$10 or 11), the Ba nucleus breaks (fissions) in to two symmetrical fragments, releasing from 
the neck region some clusters of hydrogen isotopes $^{2,3}H$ or $^4$He. This is an interesting result of the RMF(NL3) 
technique for nuclear structure physics. 

Clustering is also important for the decay of excited compound nuclei formed in nuclear reactions, where nuclei in the neck
region could have correspondance with the measured fusion-evaporation residues consisting of light particles with $Z\le$2. 
We know from the Introduction above that todate the decay of $^{116,118,122}$Ba$^*$ compound nuclei in to intermediate mass 
fragments (IMFs), and symmetric and near-symmetric fission fragments are measured 
\cite{campo88,campo91,campo98,commara2k,bonnet08}, but the fusion-evaporation cross-sections are not yet measured.

\section{Acknowledgments}
 
This work is supported in part by Council of Scientific $\&$ Industrial Research (No.03(1060)06/EMR-II), as well as the 
Department of Science and Technology, Govt. of India, project No. SR/S2/HEP-16/2005.

\end{document}